\documentclass[prb,twocolumn,superscriptaddress,amsmath,amssymb,floatfix,preprintnumbers]{revtex4}
\usepackage{amsmath}
\usepackage{graphicx}
\usepackage{dcolumn}
\usepackage{color}
\DeclareGraphicsExtensions{.png .jpg .pdf}

\usepackage{dsfont}
\begin{document}

\title{Current modulation in graphene p-n junctions with external fields}

\author{F. R. V. Ara\'ujo} \email{ronan.viana@fisica.ufc.br}
\affiliation{Departamento de F\'isica, Universidade Federal do Cear\'a, Campus do Pici, 60455-900 Fortaleza, Cear\'a, Brazil}
\affiliation{Instituto Federal de Educa\c{c}\~ao, Ci\^encia e Tecnologia do Piau\'i, Campus S\~ao Raimundo Nonato, 64670-000, S\~ao Raimundo Nonato, Piau\'i, Brazil}
\author{D. R. da Costa}\email{diego_rabelo@fisica.ufc.br}
\affiliation{Departamento de F\'isica, Universidade Federal do Cear\'a, Campus do Pici, 60455-900 Fortaleza, Cear\'a, Brazil}
\author{A. C. S. Nascimento}\email{alexandro@ifpi.edu.br}
\affiliation{Instituto Federal de Educa\c{c}\~ao, Ci\^encia e Tecnologia do Piau\'i, Campus Parna\'iba, 64211-145, Parna\'iba, Piau\'i, Brazil}
\author{J. M. Pereira Jr.}\email{pereira@fisica.ufc.br}
\affiliation{Departamento de F\'isica, Universidade Federal do Cear\'a, Campus do Pici, 60455-900 Fortaleza, Cear\'a, Brazil}

\date{\today}

\begin{abstract}
In this work we describe a proposal for a graphene-based nanostructure that modulates electric current even in the absence of a gap in the band structure. The device consists of a graphene p-n junction that acts as a Veselago lens that focuses ballistic electrons on the output lead. Applying external (electric and magnetic) fields changes the position of the output focus, reducing the transmission. Such device can be applied to low power field effect transistors, which can benefit from graphene's high electronic mobility.
\end{abstract}


\maketitle

\section{Introduction}\label{sec1}

The production of high-quality samples of graphene has allowed the investigation of charge transport in the ballistic regime at length scales much larger than in other materials \cite{katsnelson2007graphene}. This fact has permitted the observation of effects such as Klein tunneling \cite{katsnelson2006chiral,pereira2010klein, wilmart2014klein} and Fabry-P\'erot oscillations \cite{pereira2007graphene,masir2010fabry,shytov2008klein} that point towards a striking similarity between light propagation in waveguides and electronic transport in graphene. Due to Klein tunneling, \textit{i.e.} the perfect transmission through potential barriers, the confinement of charge carriers in graphene can become a challenge. That fact limits the use of graphene on logical device applications, due to the fact that one cannot in general ``turn off'' the current. Some ways to circumvent that limitation are the use of graphene nanoribbons \cite{brey2006electronic, wakabayashi2009electronic, wakabayashi2010electronic, rozhkov2011electronic, stampfer2011transport, nascimento2019electronic,son2006energy}, in which the geometry of the sample induces a lateral confinement that can create a gap in the band structure. Other possibilities involve the use of graphene bilayers, or the application of strain which can also give rise to a band gap \cite{neto2009electronic}.

It has been recently shown that an additional mechanism for controlling the propagation of electrons in graphene without the creation of a band gap can be developed in analogy with an optical counterpart, namely, phase modulation \cite{shu2018significantly}. Optical phase modulators make use of the electro-optical effect, in which a voltage can change the refractive index of a given medium. In the case of graphene, a similar effect can be obtained by means of p-n junctions \cite{milovanovic2015veselago, phong2016fermionic, cheianov2007focusing, pendry2007negative}. Another similarity between electronic transport in graphene and optics is the negative refraction of electrons incident on a p-n junction \cite{pendry2007negative, phong2016fermionic}. In photonic systems, a medium with a negative refraction index would allow the development of devices such as superlenses which can focus light beams beyond the diffraction limit.\cite{padilla2006negative} The prospect of an optical superlens was first raised by V. G. Veselago\cite{veselago1968electrodynamics}, who showed that in conditions where the electric and magnetic responses are negative, the group and phase velocities presented opposite directions. For electrons in graphene, theoretical \cite{cheianov2007focusing, cheianov2006selective, abanin2007quantized, shytov2009atomic, low2009ballistic, low2009electronic, milovanovic2013graphene, wilmart2014klein, milovanovic2014magnetic, phong2016fermionic, reijnders2017symmetry} and experimental works \cite{lee2015observation, stander2009evidence, huard2007transport,lin2015building, chen2016electron} have shown this effect, which can also be exploited to focus electron beams with high precision.

The ability to focus an electron beam on a small region of a graphene sample suggests that a Veselago lens may allow the development of a current switch that can be operated by properly applying an external (electric or magnetic) field. Thus, in this work we theoretically investigate a graphene-based device in which electrons emitted from an input lead are focused by a p-n junction on an output lead, so that the overall transmission amplitude and therefore the conductance of the device are increased. Applying an in-plane electric field or a perpendicular magnetic field acts to shift the position of the focal point, increasing reflectance and thus significantly decreasing transmission. The device bears some resemblance to the optical technique for imaging fluids known as Schlieren photography\cite{settles2006high}, in which light from a collimated sourced is focused on a knife edge that blocks half the incoming light, such that small changes in fluid density result in large variations in image contrast.

The paper is organized as follows. In Sec.~\ref{sec2} we present the theoretical framework used to describe the transport properties of the graphene-based current modulator as well as its operating characteristics. In Sec.~\ref{sec3} we discuss the numerical results and analyze it within a semiclassical picture. Finally, in Sec.~\ref{sec4} we summarize our main findings.

\section{Model}\label{sec2}

\begin{figure}[t]
\centering
\includegraphics[width = \linewidth]{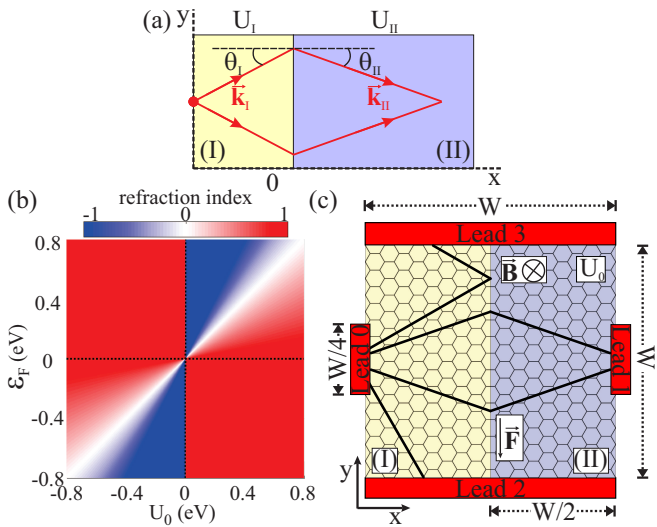}
\caption{(color online) (a) Sketch of focusing effect of electrons in graphene across a potential step due to negative refraction index. In region I (II) is applied a bias potential $U_I$ ($U_{II}$). The incident and transmitted electronic waves have momentum $\vec{k}_I$ and $\vec{k}_{II}$ and angles $\theta_I$ and $\theta_{II}$ formed with $x$-axis, respectively. (b) Refraction index, given by Eq.~(\ref{eq2}), as a function of the electrostatic potential strength $U_0$ and Fermi energy $\epsilon_F$, taking $U_I=0$ and $U_{II} = U_0$ for the bias potential in each junction region. (c) Schematic representation of the proposed current modulator. In the region II at the right-half of the sample, a bias potential with amplitude $U_0$, an in-plane electric field $\vec{F}$ and a perpendicular magnetic field are inserted. The electrons are injected into the scattering region by lead $0$ in region I and can be collected by leads $1$, $2$ or $3$. The square system length is $W$ and the lead width is assumed as $W/4$.}
\label{system}
\end{figure}

Before describing the proposed nanostructure, let us briefly recall the propagation of an electronic plane wave through a potential step in graphene in order to elucidate how a graphene p-n junction acts as focusing lens for electrons.\cite{pendry2007negative, phong2016fermionic, cheianov2007focusing} Let us consider the system shown in Fig.~\ref{system}(a) with different charge densities on regions I ($x<0$) and II ($x>0$) induced by two gates that shift the Dirac cones by $U_I$ and by $U_{II}$, respectively. An electron approaching the junction from region I reaches the interface with an incident angle $\theta_I$ and is transmitted to region II with a transmission angle $\theta_{II}$, where $\vec{k}_I$ and $\vec{k}_{II}$ are the respective wavevectors. Since the system has translational symmetry along the $y$-direction, the transverse momentum ($k_y$) is conserved at the interface, such that $|\vec{k}_{I}|\sin\theta_I = |\vec{k}_{II}|\sin\theta_{II}$. From the Dirac equation for biased graphene, we have the shifted dispersion relation $\epsilon_F(k) = s\hbar v_F k + U_i$, where $s=+/-$ correspond to electrons/holes, respectively, $i$ denotes the region index $I$ and $II$, and $v_F$ is the Fermi velocity. Connecting both equations, it implies in a similar Snell's law to ray optics where the energies here play the role of the refractive index:
\begin{eqnarray}\label{eq2}
n = \frac{\sin\theta_I}{\sin\theta_{II}}=\frac{k_{II}}{k_I}=\frac{\epsilon_F-U_{II}}{\epsilon_F-U_{I}}.
\end{eqnarray}
Note that when $\left(\epsilon_F-U_{II}\right)\left(\epsilon_F-U_{I}\right)<0$, or equivalently when $\theta_I$ and $\theta_{II}$ have opposite signs, one obtains a negative refractive index impling that the sign of the tangential momentum component of the propagating electron changes while the normal component remains the same. As a consequence, the incident electrons will converge into a focal point on region II, as it happens in Veselago lens medium\cite{pendry2007negative, phong2016fermionic, cheianov2007focusing}. The refractive index is shown in Fig.~\ref{system}(b) as a function of the electrostatic potential strength $U_0$ and Fermi energy $\epsilon_F$, for the system parameter ranges investigated along this work and taking $U_I=0$ and $U_{II} = U_0$ for the bias potential in regions I and II, respectively. It is easy to see from Eq.~(\ref{eq2}) for $U_I=0$ and $U_{II} = U_0$ that: the negative (positive) refraction index $n<0$ ($n>0$) happens when $U_0>\epsilon_F$ ($U_0<\epsilon_F$) such that the electron semiclassical trajectories are expected to (converge) diverge with the p-n junction interface acting as (convex) concave lens, while for $U_0=\epsilon_F$ one gets the $n=0$ situation where the transmission angle is zero and the electrons are perfect collimated. The blue, red and white colors denote $n<0$, $n>0$ and $n=0$ cases in Fig.~\ref{system}(b). Therefore, there is a direct analogy between propagated charge carriers through graphene p-n junction and the light focusing observed in Veselago lens due to negative refraction index medium.

Motivated by this negative refractive effect on a graphene p-n junction, we propose the nanostructure schematically illustrated in Fig.~\ref{system}(c) as a current modulator, as will be justified by the results discussed in Sec.~\ref{sec3}. For this, we investigate the transmission of these electrons through an abrupt biased graphene p-n junction, \textit{i.e.} graphene in the presence of a potential step created by electrostatic gates, and in addition to that we include an in-plane electric field and a perpendicular magnetic field to tune the electron focus and consequently to modulate the conductance. The energy spectrum for this system was analytically studied in detail in Ref.~[\onlinecite{peres2007algebraic}], the transmission probability and conductance in the absence of magnetic field for graphene p-n junction has been shown in Refs.~[\onlinecite{cheianov2007focusing,cheianov2006selective, low2009ballistic, low2009electronic,milovanovic2015veselago, wilmart2014klein, lee2015observation, reijnders2017symmetry}], and in the presence of magnetic field has been reported in Refs.~[\onlinecite{abanin2007quantized, cheianov2006selective, shytov2009atomic, stander2009evidence, milovanovic2013graphene, milovanovic2014magnetic, chen2016electron}]. Although in a more realistic experimental set-up the p-n junction has a finite width, it has been shown\cite{phong2016fermionic} that smooth graphene p-n junctions exhibit negative refraction and lensing similar to a sharp junction, as considered in the current work. 

The system consists of a square graphene sample with length $W$. The in-plane electric and magnetic fields are just applied in region II, \textit{i.e.} on the right-half of the sample ($x>0$). For the sake of simplicity, we consider only nanoribbons with armchair-type edges, although one expects the results are not qualitatively distinct from the zigzag case, since we are not concerned with edge states for the transport properties. Indeed, it has been shown in Ref.~[\onlinecite{milovanovic2015veselago}] that zigzag interface and armchair interface graphene p-n junction exhibits qualitatively similar results. Moreover, the actual sample studied throughout the paper is large enough such that the qualitative behaviors here reported and the proof-of-concept of the proposed system as a current modulator, which are the main goals of this work, would still hold, since they are based on more fundamental physical properties regarding the negative refractive effect of the proposed structure, as we discuss bellow.

Four ballistic leads are added to the scattering region with two of them parallel to the $x$-axis (leads $2$ and $3$ that cover the entire top and bottom boundaries), while the other two leads are perpendicular to that direction (leads $0$ and $1$ that are narrow with respect to the total system size in order to represent a point source and a focal point, respectively). In this configuration, the electrons are injected into the scattering region by lead $0$ (left side of the sample) and can be collected by leads $1$, $2$ or $3$, as shown in Fig.~\ref{system}(c). Leads $2$ and $3$ are included to prevent interference from electrons that do not reach the region II of Fig.~\ref{system}(c) and due to unintended reflections from the edges of the sample.

In order to show that the proposed system works as a current modulator, we shall calculate and discuss the probability current density and conductance as a function of several parameters, such as: Fermi energy, potential step height, magnetic and electric field amplitudes, and different widths of the scattering region. Our theoretical framework is based on the Landauer-B\"uttiker formalism where the transport properties are computed within the wave function approach\cite{groth2014kwant} and by using the tight-binding model within the nearest-neighbor approximation to describe the charge carriers in graphene. The corresponding Hamiltonian can be written as 
\begin{equation}\label{TBHamiltonian}
H_{TB} = \sum_{i}(\epsilon_i+U_i+F_i)c^{\dagger}_ic_i + \sum_{i\neq j}(\tau_{ij} c_i^{\dagger}c_j + h.c),
\end{equation}
where $c_{i}$ ($c_{i}^{\dagger}$) annihilates (creates) an electron in site $i$ with on-site energy $\epsilon_i$. $\tau_{ij} = t =-2.8$ eV is the nearest-neighbor hopping parameter between the atoms in the $A$ and $B$ sublattices. $U_i$ and $F_i$ are on-site potentials that are used here to simulate the p-n junction and to apply the in-plane electric field, respectively. The gate potential $U_i$ consists of a single step at $x = 0$: $U_i=U_0\Theta(x)$, where $\Theta(x)$ is the Heaviside step function and $U_0$ is the potential height. The in-plane electric field is applied along the $y$-direction, \textit{i.e.} $\vec{F}=(0,-F_y,0)$, and  is perpendicular to the propagation direction between leads $0$ and $1$. The effect of an external magnetic field is incorporated in the tight-binding model via the Peierls substitution, as $\tau_{ij} \rightarrow \tau_{ij} \exp\left[i\frac{e}{\hbar}\int_j^i \vec{A}\cdot d\vec{l}\right]$, where $\vec{A}$ is the vector potential associated with an external magnetic field $\vec{B}$, that we assume here to be perpendicular to the graphene flake, $\vec{B}= B\hat{z}$.

All the transport calculations presented in this work were performed using the KWANT code, which is a free (open source) Python package for numerical calculations on tight-binding models. \cite{groth2014kwant}

\section{Results}\label{sec3}

\begin{figure}[t]
\centering
\includegraphics[width = \linewidth]{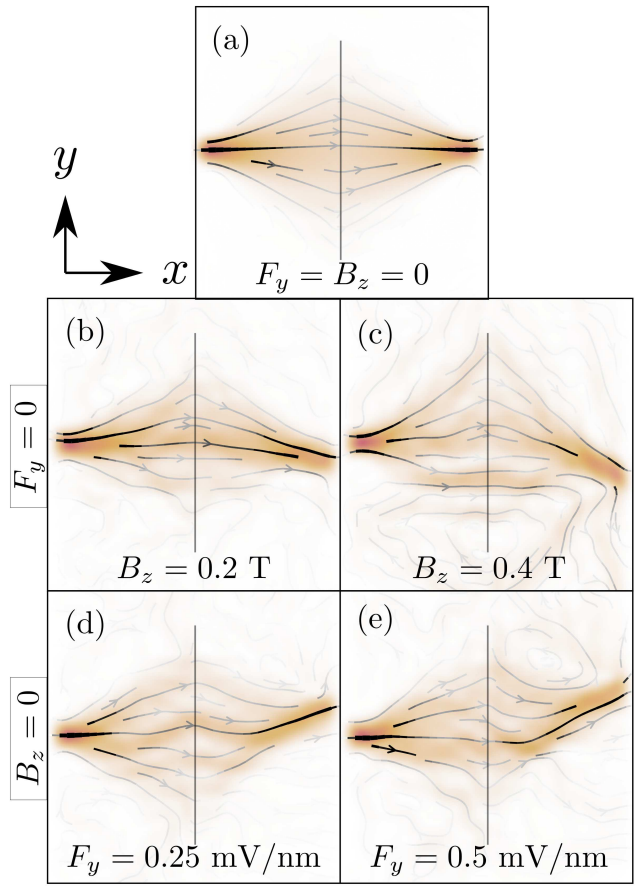}
\caption{(color online) Probability current densities for the system shown in Fig.~\ref{system}(b) for (a) the non-perturbed case with $B_z=0$ and $F_y=0$, and under the effect of perpendicular magnetic field (with (b) $B_z=0.2$ T and (c) $B_z=0.4$ T) and in-plane electric field (with (d) $F_y = 0.25$ mV/nm and (e) $F_y = 0.5$ mV/nm). Plots are made using a very narrow input and a large output leads in order to mimic a source point and to avoid backscattering when the focus position changes.}
\label{current}
\end{figure}

In order to verify that the system works as a Veselago lens, we first calculate the probability current density as shown in Fig.~\ref{current} for the graphene p-n junction system sketched in Fig.~\ref{system}(c) and show that the focal point moves by applying external fields: perpendicular magnetic field in Figs.~\ref{current}(b)-\ref{current}(c) and in-plane electric field in Figs.~\ref{current}(d)-\ref{current}(e). Without loss of generality, we consider a symmetric junction with respect to the scattering region size such that the injected and collecting leads are placed $W/2$ away from the interface at $x=0$, and the Fermi energy as $\epsilon_F = U_0/2 = 0.4$ eV. We assume the injected lead to be narrow enough in order to mimic a source point, whereas the collecting lead here for these results was made larger to avoid backscattering and thus a misleading understanding of the lensing process. For the non-perturbed case [Fig.~\ref{current}(a) for $F=0$ and $B=0$], the propagated wave is focused symmetrically such that the focus spot is at the same height $y$ as the source lead. By applying a perpendicular magnetic field the current density vectors are seen to be deflected (downwards since $\vec{B}$ points into the page) due to the Lorentz force, shifting the focal point position, as depicted in Figs.~\ref{current}(b) and \ref{current}(c) for magnetic field amplitudes $B=0.2$ T and $B=0.4$ T, respectively. By comparing Figs.~\ref{current}(b) and \ref{current}(c), one can notice the focal position shift is larger the higher the magnetic field amplitude. This can be easily understood by the following semiclassical picture: from Lorentz force $m\vec{a} = -e \vec{v}\times\vec{B}$, where $e$ is the elementary charge, $\vec{a}$ and $\vec{v}$ are the acceleration and speed of electron, respectively, and knowing that the cyclotron effective mass depends on the band structure via the derivative of this area in energy\cite{abdullah2019electron,ariel2013electron} such that for an isotropic energy spectrum one has $m=\hbar^2/(2\pi)(d^2 A(\epsilon)/d\epsilon^2)$, where $A(\epsilon)$ denotes the $k$-space area enclosed by a constant energy contour $\epsilon$, one can find the cyclotron radius as $r_c = \left|\epsilon/(ev_F B) \right|$. This shows that the cyclotron radius is inversely proportional to the magnetic field and therefore the larger $|B|$ the smaller is $r_c$ and consequently  the ($x$,$y$)-coordinates of the focal point vary more. A similar effect can be achieved when an in-plane electric field is applied in the region II of the system, as shown in Figs.~\ref{current}(d) and \ref{current}(e) for $\vec{F}=(0,-F_y,0)$ with $F_y = 0.25$ mV/nm and $F_y = 0.5$ mV/nm, respectively, such that the negative charge carriers are pushed upwards and the focal point moved up along $y$-direction.

\begin{figure}[t]
\centering
\includegraphics[width = \linewidth]{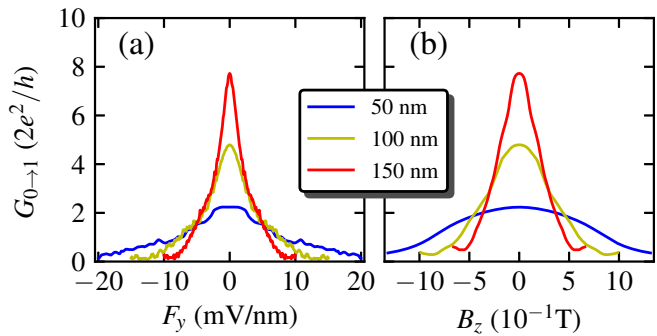}
\caption{(color online) Conductance as a function of the electric (a) and magnetic (b) field amplitudes between leads $0$ and $1$ for sample width (blue curve) $W=50$ nm, (yellow curve) $W=100$ nm, and (red curve) $W=150$ nm. It is taken $\epsilon_F=U_0/2=0.4$ eV, and for panel (a) $B_Z=0$ and for panel (b) $F_y=0$.}
\label{fields}
\end{figure}

\begin{figure*}[b]
\centering
\includegraphics[width = 0.85\linewidth]{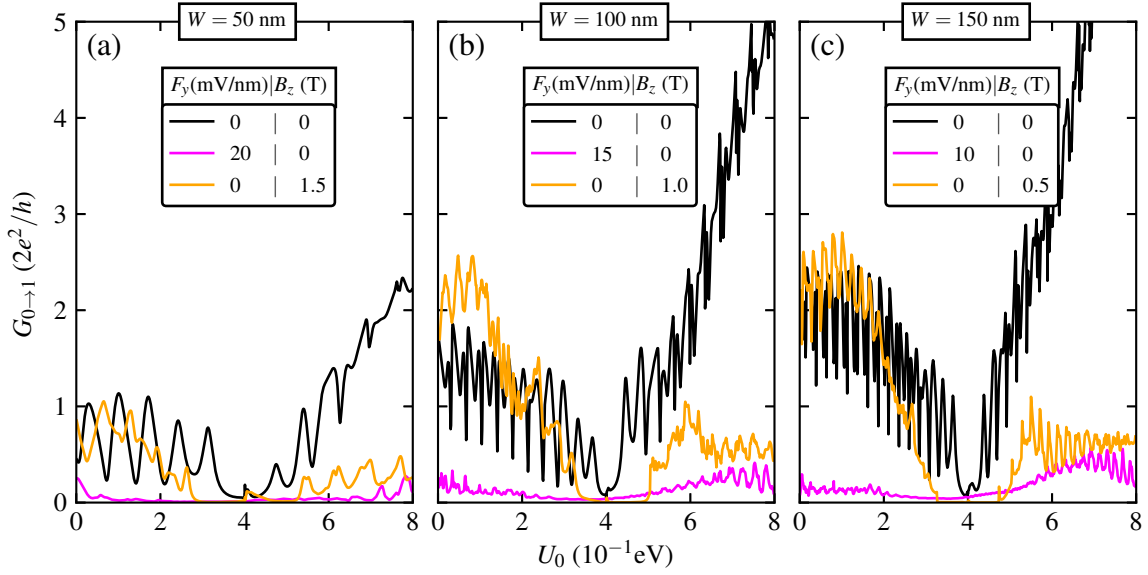}
\caption{(color online) Conductance between leads $0$ and $1$ as a function of the potential step height $U_0$ in the absence of any external field (black curve for $B_z=0$ and $F_y=0$), and in the presence of an in-plane electric field (magenta curve for $F_y\neq 0$) and an out-of-plane magnetic field (orange curve for $B_z\neq 0$). Panels (a), (b) and (c) correspond to sample width $W=50$ nm, $W=100$ nm, and $W=150$ nm, respectively. It is taken $\epsilon_F=0.4$ eV.}
\label{potential}
\end{figure*}

In both external electric and magnetic field cases discussed in Fig.~\ref{current}, as high the amplitude of the external field more further away the focal point is from the output lead and consequently a decreasing in the conductance is expected. In order to quantify how the conductance changes due to the external fields, we perform numerical calculations of the conductance between leads $0$ and $1$ as a function the electric [Fig.~\ref{fields}(a)] and magnetic [Fig.~\ref{fields}(b)] field amplitudes by considering the symmetric Veselago lens case that means the focal spot is placed at the same distance from the p-n interface as the injector. It corresponds to $\theta_I = -\theta_{II}$ case, or equivalently the situation where the Fermi energy must be half value of the potential step height, \textit{i.e.} $\epsilon_F=U_0/2$, that means $n=-1$ in Eq.~(\ref{eq2}). This symmetrical p-n junction implies that the number of electrons injected by lead $0$ and captured by lead $1$ is the maximum that can be achieved for this setup. The results shown in Fig.~\ref{fields} were obtained for three different sample widths $W$, viz. $50$, $100$ and $150$ nm, and taking $\epsilon_F=U_0/2=0.4$ eV.  

\begin{figure*}[b]
\centering
\includegraphics[width = 0.85\linewidth]{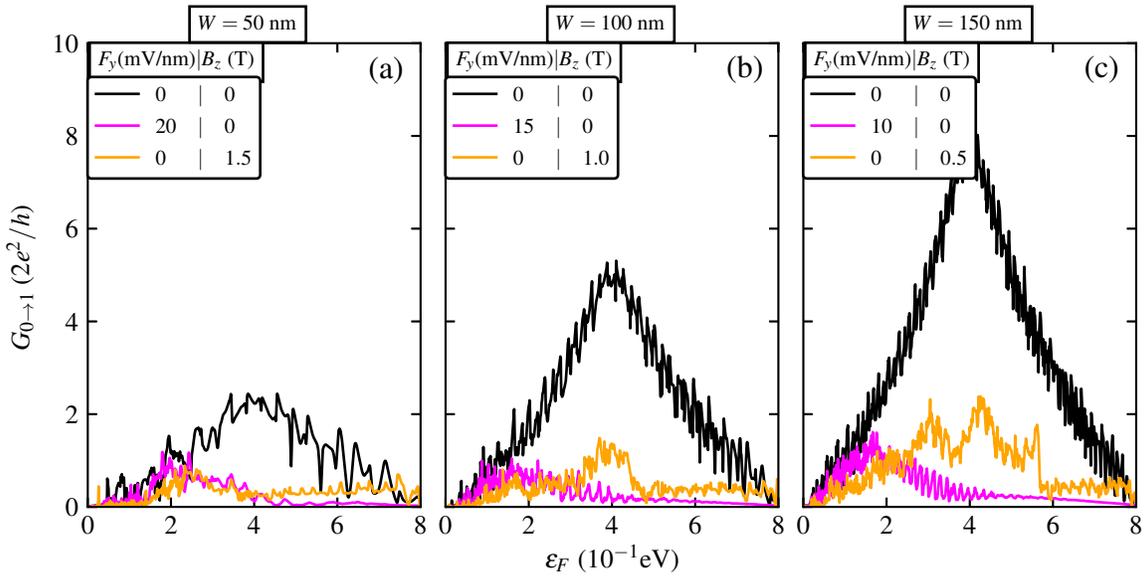}
\caption{(color online) Conductance between leads $0$ and $1$ as a function of the Fermi energy $\epsilon_F$ in the absence of any external field (black curve for $B_z=0$ and $F_y=0$), and in the presence of an in-plane electric field (magenta curve for $F_y\neq 0$) and an out-of-plane magnetic field (orange curve for $B_z\neq 0$). Panels (a), (b) and (c) correspond to sample width $W=50$ nm, $W=100$ nm, and $W=150$ nm, respectively. It is taken $U_0=0.8$ eV.}
\label{energy}
\end{figure*}

In Fig.~\ref{fields}, one can notice that the conductance between leads $0$ and $1$ is significantly reduced by the application of the external in-plane electric [Fig.~\ref{fields}(a)] and perpendicular magnetic [Fig.~\ref{fields}(b)] fields, as expected based on the fact that focal spot in region II of the sample is shifted in the $y$-direction a way from the collector (lead $1$). As a consequence, the electron beams are then scattered by the right boundary of the graphene p-n junction and collected by the bottom and top leads $2$ and $3$, respectively, leading to a current reduction between leads $0$ and $1$. Furthermore, one can verify from Fig.~\ref{fields} that the external field required to control the conductance depends on the system width $W$, and the larger the width considered the smaller the external field needed to obtain a significant change on $G_{0\rightarrow 1}$, as for instance in Fig.~\ref{fields}(a) $G_{0\rightarrow 1}\approx 0$ for $W=150$ nm with $F_y\approx 10$ mV/mm, for $W=100$ nm with $F_y\approx 15$ mV/mm, and for $W=50$ nm with $F_y\approx 20$ mV/mm. Let us now understand this $W$ dependence on the conductance. Note that: (i) for the non-perturbed case with $B_z = 0$ and $F_y = 0$, the conductance is larger the wider the system, \textit{i.e.} $G_{0\rightarrow 1}(W=150\mbox{ nm})>G_{0\rightarrow 1}(W=100\mbox{ nm})>G_{0\rightarrow 1}(W=50\mbox{ nm})$; (ii) due to the finite system along $y$-direction the sample can be seen as a large armchair nanoribbon with ribbon width $W$ which in turn resembles a potential well in $y$-direction; and (iii) for graphene nanoribbons,\cite{brey2006electronic, wakabayashi2009electronic, wakabayashi2010electronic, rozhkov2011electronic, stampfer2011transport} as the width $W$ increases there are more localized states for a fixed energy range and the energy levels become closer. Therefore in general for larger systems there are more transverse electronic modes available to contribute to transport and thus the conductance is larger the wider the width $W$. Although the qualitative behavior of conductance as a function of external electric and magnetic fields are similar, in the case of non-zero magnetic field the minimal conductance obtained is larger than the minimal conductance for the electric field case. For instance the minimal conductance in units of $2e^2/\hbar$ for $W=150$ nm and $W=100$ nm in the presence of electric field (panel (a)) are $0.113$ and $0.098$, respectively, while for non-zero magnetic field (panel(b)) are $0.483$ and $0.284$, respectively.

Let us now analyze the influence of the potential step height on the conductance between leads $0$ and $1$. Fig.~\ref{potential} shows the conductance as function of $U_0$ for the same three different sample widths described above: [Fig.~\ref{potential}(a)] $W = 50$ nm, [Fig.~\ref{potential}(b)] $W = 100$ nm, and [Fig.~\ref{potential}(c)] $W = 150$ nm, in the absence of any external field (black curve for $B_z=0$ and $F_y=0$), and in the presence of an in-plane electric field (magenta curve for $F_y\neq 0$) and an out-of-plane magnetic field (orange curve for $B_z\neq 0$). In particular, for non-zero external field it was assumed the approximated values of the electric and magnetic field amplitudes associated to the minimal conductance shown in Fig.~\ref{fields}, such as, e.g. for finite electric (magnetic) field and for $W=50$ nm the value $F_y=20$ mV/nm ($B_z = 1.5$ T), for $W=100$ we use $F_y=15$ mV/nm ($B_z = 1.0$ T) and for $W=150$ nm, $F_y=10$ mV/nm ($B_z = 0.5$ T). One can notice from Fig.~\ref{potential} that the conductance exhibits an asymmetric behavior with respect to ($U_0=\epsilon_F$)-axis, \textit{i.e.} there are two different trends: for $U_0<\epsilon_F$ and $U_0>\epsilon_F$, where it was taken $\epsilon_F=0.4$ eV. From Eq.~(\ref{eq2}) and its analysis discussed in Sec.~\ref{sec2}, the nature of the two different conductance regimes is related to the different signs of the refraction index, being positive and negative for $U_0<\epsilon_F$ and $U_0>\epsilon_F$, respectively. In the negative (positive) refraction index regime, one can observe an increase (a decrease) of conductance with the gate potential increase in Fig.~\ref{potential}. This, in turn, is related to the fact that the value of the refraction index will dictate the focal spot position, since depending on its value the electron beams will be perfectly collimated, or converge or diverge after reaching the interface, and therefore a portion of the injected electrons will be captured or not by the output lead and also determinating the conductance amplitude. From Eq.~(\ref{eq2}) and by a simple geometry analisis of electronic trajectory sketched in Fig.~\ref{system},\cite{milovanovic2015veselago} one can predict the $x$-position of the focal spot, as 
\begin{equation}\label{eq.focus}
x_{focal} = \frac{W}{2} \left|\frac{\tan\theta_I}{\tan\theta_{II}} \right|,
\end{equation}
where the transmitted angle for investigated potential step with $U_I = 0$ and $U_{II} = U_0$ is given by
\begin{equation}\label{eq.theta}
\theta_{II} = \arcsin\left[\left(\frac{\epsilon_F}{\epsilon_F-U_0}\right)\sin\theta_I\right].
\end{equation}
As expected for the symmetric case, one has $\theta_I = -\theta_{II}$ and thus one can find $x_{focal} = \frac{W}{2}$. From Eqs.~(\ref{eq.focus}) and (\ref{eq.theta}), one can realize that: the position of the focal spot depends on the gate potential value, the Fermi energy and the incident angle; and the transmission increases as $x_{focal}$ approaches the output lead. Moreover, once that we assumed a symmetrical graphene p-n junction, the maximum $G_{0\rightarrow 1}$ value in the absence of any external field (black curves in Fig.~\ref{potential}) is achieved for $U_0 \approx 2\epsilon_F = 0.8$ eV, \textit{i.e.} for $n=-1$, while for the range $\epsilon_F< U_0 < 2\epsilon_F$ one has $-1< n < 0$ with the focal $x$-position not being diametrically opposite to input lead but it will be tending to the symmetric position as $n$ tends to $-1$, and thus explaining the $G_{0\rightarrow 1}$ increasing up to its maximum value at $U_0 \approx 2\epsilon_F$.

In addition, from Fig.~\ref{potential}, one can note that although there is a significant reduction in conductance with the application of the in-plane electric field (magenta curves), with the application of the perpendicular magnetic field (orange curves) there is a less significant reduction. Most importantly for our purpose of an efficient device operating on the negative refraction index regime, is that regardless of the sample width or even the kind of the applied external field (electric or magnetic) the conductance  $G_{0\rightarrow 1}$ is reduced. Although not shown, it is easy to see that as the conductance between leads $0$ and $1$ increases, the conductance between leads $0$ and $2$ and between leads $0$ and $3$ decreases proportionally. It is important to emphasize that the interplay between the external magnetic field in Region II ($x>0$) and the bias gate induced by the p-n junction will favor a high conductance for a certain small energy range when $U_0<\epsilon_F$ as a consequence of a maximized focusing of the divergent beam due to $n>0$ and the circular orbit due to $B$ field, that brings the focal point closer to the output spot.\cite{prabhakar2019valley} Such behavior can be seen by orange curves in Figs.~\ref{potential}(a), \ref{potential}(b) and \ref{potential}(c) for $U_0 \in [0.05,0.1]$ eV that exhibits conductance peaks higher than those ones for null magnetic field case.

Next, we study the dependence of the conductance on the Fermi energy for a fixed potential step height $U_0=0.8$ eV and under the influence of external fields regarding three different sample widths. The results are shown in Fig.~\ref{energy} for the same parameters as in Fig.~\ref{potential}. For $B_z=0$ and $F_y=0$ (black curves), a pronounced peak arises in the conductance at the same position $\epsilon_F=U_0/2=0.4$ eV regardless of the sample width (see the same behavior in three panels \ref{energy}(a)-\ref{energy}(c)). This is a direct consequence of the negative refraction induced by the Veselago lens property in graphene p-n junction. Let us examine the symmetric nature of the conductance with respect to the ($\epsilon_F=U_0/2$)-axis. In the investigated Fermi energy range $0\leq \epsilon_F\leq U_0$ in the plots of Fig.~\ref{energy}, one has by Eq.~(\ref{eq2}) that $n<0$. One can split the energy range as follows: for $0\leq \epsilon_F \leq U_0/2$, the refractive index is $n\leq -1$, whereas for $U_0/2 < \epsilon_F \leq U_0$, one has that $-1<n\leq 0$. It is interesting to mention that for $n<-1$ and $-1<n\leq 0$ the semiclassical electronic trajectories are caustics, which are the envelope of the classical trajectories, with cusp points shifted in $x$-direction for left and the right with respect to the symmetric case where $n=-1$.\cite{reijnders2017symmetry} Thus, based on this classical picture, it is evident that the displacement of the focal spot in $x$-direction will cause a reduction in the conductance. As discussed previously, the maximum conductance is obtained in the symmetric situation when $\epsilon_F=U_0/2$ or equivalently when $\theta_I=-\theta_{II}$ and $n=-1$. Thus, for lower or higher Fermi energies than $U_0/2$ the conductance is lower than its maximum value. Other relevant case to analyze is when $\epsilon=U_0$. By replacing $\epsilon=U_0$ in Eq.~(\ref{eq2}), one obtains that $\theta_{II}=0$, that means the electron beams are perfected collimated in region II. If instead of a finite focal spot, one considers a focal point for the output lead, thus, in this case where $\epsilon=U_0$, just the electrons with normal incidence would be captured by the output lead, explaining the lowest conductance value at $\epsilon=0.8$ eV in Fig.~\ref{energy}. Based on these statements, it suggests that the conductance curve should decrease between the symmetric situation ($\epsilon_F=U_0/2$) that is a maximum point and the perfected collimation situation ($\epsilon=U_0$) that corresponds to a minimum. 

With the application of an in-plane electric field (magenta curves in Fig.~\ref{energy}), the conductance is strongly reduced and its maximum is pushed to the energetic range $0\leq \epsilon_F \leq U_0/2$, while the conductance is almost zero and unchanged by increasing the Fermi energy within the range $U_0/2 < \epsilon_F \leq U_0$. For instance, in Fig.~\ref{energy}(c) for $W=150$ nm, the conductance is reduced of approximately ten times as compared to the zero in-plane electric field case. On the other hand, by applying a perpendicular magnetic field (orange curves in Fig.~\ref{energy}) there is a slight reduction in conductance but not so pronounced as for non-zero electric field case. This is in agreement with the discussion made in Fig.~\ref{potential} about the influence and robustness of the electric and magnetic field on the conductance. Moreover, by comparing the orange curves in Figs.~\ref{energy}(a), \ref{energy}(b) and \ref{energy}(c), one can see that the region of the pronounced conductance reduction goes to higher energies for larger samples. This is due to the fact that the thinner the nanoribbon, fewer lower energy states will have an orbit that fits in the sample and thus be deflected towards the focus point, since in the presence of a magnetic field the electron beam is bent with a cyclotron radius that is (in)directly proportional to the Fermi energy (the $B$ field magnitude). On the other hand, the electronic orbits for high energy values will be reflected at the edges and interfere themselves, and thus resulting in a reduction of the conductance $G_{0\rightarrow 1}$.

\section{Conclusion}\label{sec4}

We proposed a current modulator-like device model based on the Veselago lensing effect in graphene p-n junction. The operating principle of this device is connected to the fact that Dirac electrons passing through graphene p-n junction at specific energy are transmitted with a negative angle and thus converge on the other sample side at the focal point. This is due to the negative refraction index merged from the energy difference ratio between the Fermi energy and bias potential in the two regions of this junction and therefore the optic-like Dirac electron behavior in a graphene p-n junction is analog of a Veselago lens. We demonstrated that an in-plane electric field or an out-of-plane magnetic field move the electronic focal spot further way from the output lead and consequently tune the current transmission between the input and output leads. For the proof-of-concept that the proposed device can work as a current modulator and for its transport properties quantification, we investigated the behavior of the probability current density and conductance, by using the Landauer-B\"uttiker formalism within the tight-binding approach, as a function of the electric and magnetic field amplitudes, the Fermi energy, the system size, and the potential step height. Our findings show that the application of the external fields to this system can reduce significantly its conductance even for low power fields.   Finally, we hope that our results and the proposed nanostructure will prove useful for designing graphene-based current modulator like optical devices that works even in the absence of a gap in the graphene band structure and in low power field regime.

\acknowledgments  This work has been financially supported by CNPq, through the PRONEX/FUNCAP and PQ programs.

\bibliographystyle{apsrev}
\bibliography{ref}

\end{document}